\documentclass[aps,prl,preprint,groupedaddress]{revtex4}
\usepackage{graphicx}
\begin{document}

\preprint{TIT/HEP--523}
\preprint{hep-th/0405129}

\title{
All Exact Solutions of a 1/4 Bogomol'nyi-Prasad-Sommerfield Equation
}

\author{Youichi~Isozumi}
\author{Muneto~Nitta}
\author{Keisuke~Ohashi}
\author{Norisuke~Sakai}
\affiliation{ Department of Physics, Tokyo Institute of 
Technology,
Tokyo 152-8551, JAPAN }


\date{\today}

\begin{abstract}
We obtain 
all possible solutions of a $1/4$ 
Bogomol'nyi-Prasad-Sommerfield 
equation 
exactly, containing configurations made of 
walls, vortices and monopoles in the Higgs phase. 
We use supersymmetric 
$U(N_{\rm C})$ gauge theories with eight supercharges 
with $N_{\rm F}$ fundamental hypermultiplets 
in the strong coupling limit. 
The moduli space for the composite solitons is found to be 
the space of 
all holomorphic maps from a complex plane to 
the wall moduli space found recently, 
the deformed complex Grassmann manifold. 
Monopoles in the Higgs phase are also found in $U(1)$ gauge theory. 
\end{abstract}


\maketitle

Dirichlet branes (D-branes) are 
Bogomol'nyi-Prasad-Sommerfield (BPS) states 
preserving a fraction of supersymmetry (SUSY) and 
have played a key role 
in non-perturbative analysis 
in string theory~\cite{Po}. 
(D-)strings ending on a D-brane have been realized 
in the effective field theory on the 
D-brane~\cite{BIon}. 
A $1/2$ BPS composite soliton of vortex (string) ending 
on a wall called a D-brane soliton has been constructed 
in a SUSY nonlinear sigma model (NLSM) with eight 
SUSY~\cite{GPTT}, which can be interpreted as 
SUSY $U(1)$ gauge theory~\cite{SY1} 
in the strong gauge coupling limit. 
Solitons such as a wall junction, 
a string intersection were constructed 
both in NLSMs~\cite{DW1} and gauge theories~\cite{DW2}
with eight SUSY. 
Recently a new $1/4$ BPS equation has been 
obtained admitting a monopole in Higgs phase 
as a kink on vortices~\cite{mono-Higgs}. 
The equation turns out to admit domain walls also 
and therefore one expects that this BPS equation 
allows interesting brane configurations made of 
these three kinds of solitons. 
In this Letter, we {\it exactly} give 
{\it all possible solutions} of the 
1/4 BPS equations 
in the SUSY $U(N_{\rm C})$ gauge theory with 
$N_{\rm F} (>N_{\rm C})$ fundamental hypermultiplets 
in the strong gauge coupling limit, 
including composite solitons made of walls, vortices 
and monopoles. 
This we find the complete moduli space 
for these solutions. 
To the best of our knowledge this is the first 
example with completely determined moduli space 
for composite solitons. 
Our results hopefully open up a new research direction 
to classify and exhaust all the BPS equations, 
their solutions and their moduli space.

We consider a five-dimensional SUSY model with minimal 
kinetic terms for vector and hypermultiplets whose 
physical bosonic fields are ($W_M, \Sigma$) and $H^{irA}$, 
respectively. 
The $A$-th hypermultiplet mass, the Fayet-Illiopoulos 
parameters, and 
a common gauge coupling constant for $U(N_{\rm C})$ 
are denoted as $m_A$, 
$c_a,\,(a=1,2,3)$, and $g$. 
After eliminating auxiliary fields, 
the bosonic part of our Lagrangian 
with the scalar potential $V$ 
reads 
\begin{eqnarray}
{\cal L} &\!\!\!=&\!\!\! 
-\frac{1}{2g^2}{\rm Tr}(F_{MN}(W)F^{MN}(W))
 +\frac{1}{g^2}{\rm Tr}({\cal D}^M \Sigma {\cal D}_M \Sigma) 
\nonumber\\
&\!\!\!&\!\!\!
+ {\cal D}_M H_{irA}^\dagger  {\cal D}^M H^{irA} -V,
\label{fundamental-Lag2}
\end{eqnarray}
\begin{eqnarray}
V
&
=
& 
\frac{g^2}{4}{\rm Tr}\left[
\left((\sigma _a)^j{}_iH^iH_j^\dagger 
-c_a{\bf 1}_{N_{\rm C}}
\right)^2\right] 
\nonumber\\&&\qquad {}
+H_{irA}^\dagger [(\Sigma -m_A)^2]^r{}_s H^{isA}
.
\label{eq:potential}
\end{eqnarray}
Covariant derivatives are 
${\cal D}_M \Sigma = \partial_M \Sigma + i[ W_M , \Sigma ]$, 
${\cal D}_M H^{irA}
=(\partial_M \delta_s^r + i(W_M)^r{}_s)H^{isA}$, 
and the gauge field strength is 
$F_{MN}(W)=-i[{\cal D}_M , {\cal D}_N]
$. 
We assume non-degenerate mass 
and $m_A > m_{A+1}$ for all $A$.  
The $SU(2)_R$ allows us to choose 
$c_a =(0,\ 0,\ c)$ with $c>0$. 

Let us obtain the 1/4 BPS equations for combined solitons 
of walls, vortices and monopoles. 
A wall preserves half of the 
eight supercharges 
defined by 
$\gamma ^3(i\sigma _3)^i{}_j\varepsilon ^j
=\varepsilon ^i$~\cite{INOS}.
We can obtain vortex preserving a different half 
defined by another projection 
$\gamma ^{12}(i\sigma _3)^i{}_j\varepsilon ^j=-\varepsilon ^i$. 
Combining them together, 
we preserve $1/4$ 
of supercharges. 
The resulting Killing spinor automatically 
satisfies the third projection 
$\gamma ^{123}\varepsilon ^i=\varepsilon ^i$, 
which allows a monopole. 
We assume the solutions depend on $x^1, x^2, x^3$ 
(co-dimension three) and assume Poincar\'e invariance 
in $x^0,x^4$. 
Then we obtain $W_0=W_4=0$. 
We have proved $H^2=0$ for 1/2 BPS saturated 
(pararell) walls in the case of non-degenerate 
masses~\cite{INOS}. 
A similar argument can be applied for the 1/4 BPS solutions 
to obtain $H^2=0$.  
Requiring the SUSY transformation of fermions to vanish 
along the above $1/4$ SUSY directions, 
a set of 1/4 BPS equations 
is obtained in the matrix notation as~\cite{mono-Higgs} 
\begin{eqnarray}
0&=&{\cal D}_3 H^1 +\Sigma H^1 -H M,\label{1/4BPSeq-1}\\
0&=&{\cal D}_1 H^1 +i{\cal D}_2 H^1,\label{1/4BPSeq-2}\\
0&=&{}^*F_1-{\cal D}_1 \Sigma,\quad 0={}^*F_{2}-{\cal D}_2 
\Sigma,\label{1/4BPSeq-4}\\
0&=&{}^*F_3-{\cal D}_3 \Sigma+{g^2\over 2}
\left(c{\bf 1}_{N_{\rm C}}-H^1H^1{}^\dagger\right) ,
\label{1/4BPSeq-5}
\end{eqnarray}
where $(M)^A{}_B\equiv m_A\delta ^A{}_B$ and 
${}^*F_m\equiv \frac12\epsilon _{mnl}F^{nl}(W)$ 
with $m=1,2,3$. 

We obtain 
the BPS bound of the energy density 
$\cal E$ as 
${\cal E}\geq t_{\rm w}+t_{\rm v}+t_{\rm m}+\partial _mJ_m$ 
with $t_{\rm w},\,t_{\rm v}$ and $t_{\rm m}$ 
the energy densities for walls, vortices and monopoles, 
and the correction term $J_m$, which does not contribute 
for individual walls and vortices 
\begin{eqnarray}
 t_{\rm w}&=&c\partial _3 {\rm Tr}\Sigma,\quad 
 t_{\rm v}=-c{\rm Tr}{}^* \!  F_3,\nonumber\\ 
 t_{\rm m}&=&\frac{2}{g^2}\partial _m {\rm Tr}(\Sigma {}^* \! F_m), 
 \label{energy-den}
\end{eqnarray}
\begin{eqnarray}
 J_1&=&{\rm Re}\left(-i{\rm Tr}(H^1{}^\dagger {\cal D}_2H^1)\right),\, 
J_2={\rm Re}\left(i{\rm Tr}(H^1{}^\dagger {\cal D}_1H^1)\right),\nonumber\\
J_3&=&-{\rm Tr}(H^1{}^\dagger (\Sigma -M)H^1).
\end{eqnarray}
Let us note that the magnetic flux from our monopole is 
measured in terms of the dual field strength multiplied 
(projected) by the Higgs field $\Sigma {}^* \! F_m$, 
as is usual to obtain the $U(1)$ field strength for 
the monopole in the Higgs phase~\cite{mono-Higgs}. 

Let us construct solutions for 
the BPS eqs.(\ref{1/4BPSeq-1})-(\ref{1/4BPSeq-5}), 
following the method to obtain complete solutions 
of non-Abelian walls~\cite{INOS}. 
It is crucial to observe that eq.(\ref{1/4BPSeq-4}) 
guarantees the integrability condition 
$[{\cal D}_1+i{\cal D}_2,\,{\cal D}_3+\Sigma]=[\partial_1+i \partial_2,\,\partial_3]=0$. 
Therefore we can introduce an $N_{\rm C}\times N_{\rm C}$ 
invertible complex matrix function 
$S(x^m)\in GL(N_{\rm C},\mathbf{C})$ 
defined by 
\begin{eqnarray}
\Sigma + iW_3 &\equiv& S^{-1}\partial_3 S,\quad 
\left(({\cal D}_3+\Sigma)S^{-1}=0\right),
\label{def-S-3} \\
W_1+iW_2&\equiv &-2iS^{-1}\bar \partial S,\, 
\left(({\cal D}_1+i{\cal D}_2)S^{-1}=0\right) , 
\quad \label{def-S-bar}
\end{eqnarray}
where $z \equiv x^1 + i x^2$, and 
$\bar \partial\equiv \partial/\partial z^*$. 
With (\ref{def-S-3}) and (\ref{def-S-bar}), 
Eq.~(\ref{1/4BPSeq-4}) is automatically satisfied, and 
Eqs. (\ref{1/4BPSeq-1}) and (\ref{1/4BPSeq-2}) 
are easily solved without any assumptions by 
\begin{eqnarray}
 H^1 = S^{-1}(z,z^*, x^3)H_0(z) e^{M x^3}.
\label{sol-H}
\end{eqnarray}
Here $H_0(z)$ is an arbitrary $N_{\rm C} \times N_{\rm F}$ 
matrix whose elements are arbitrary holomorphic functions of $z$, 
which we call  ``moduli matrix''.
Let us define an $N_{\rm C}\times N_{\rm C}$ 
Hermitian matrix $\Omega \equiv SS^\dagger $, 
invariant under the $U(N_{\rm C})$ gauge transformations 
$S\rightarrow SU^\dagger $ with $U \in U(N_{\rm C})$. 
The remaining BPS eq.~(\ref{1/4BPSeq-5}) 
can be rewritten in terms of this matrix $\Omega $ 
and the moduli matrix $H_0$ as
\begin{eqnarray}
&&4\partial \bar \partial \Omega 
-4(\partial \Omega )\Omega ^{-1}
(\bar \partial \Omega )
+\partial _3^2\Omega -(\partial _3\Omega )
\Omega ^{-1}(\partial _3\Omega )\nonumber \\
&&= g^2 
\left(c\, \Omega - H_0 \,e^{2My} H_0{}^\dagger 
\right). \label{diff-eq-S}
\end{eqnarray}
Eqs.(\ref{def-S-3})--(\ref{diff-eq-S}) 
determine a map 
from our moduli matrix $H_0(z)$ 
to all possible $1/4$ BPS solutions in three-dimensional 
configuration space. 
Let us stress that our moduli matrix $H_0(z)$ should be 
the full initial data for this map. 
The nonlinear partial differential eq.(\ref{diff-eq-S}) 
should determine $\Omega$ in terms of the moduli matrix 
$H_0$, with the aid of appropriate boundary conditions. 
From the experience of walls, we expect that there is 
no more integration constants for $\Omega$ \cite{INOS}. 
This expectation is explicitly borne out in the explicit 
solution at infinite gauge coupling as we show below. 

The first two energy densities in (\ref{energy-den}) 
can be combined in terms of $\Omega $ as 
\begin{eqnarray}
t_{\rm wv}\equiv t_{\rm w}+t_{\rm v}={c\over 2}\,
\partial _m\partial _m\log{\rm det}\Omega .  
\end{eqnarray}
By using the BPS equations, 
we find the correction term of the energy density as 
\begin{eqnarray}
 \partial _mJ_m=-{1\over 2g^2}(\partial _m\partial _m)^2
\log{\rm det}\Omega ,
\end{eqnarray}
which can be neglected if gauge coupling 
is large enough. 

Though Eq.~(\ref{diff-eq-S}) is difficult to solve 
explicitly for finite gauge couplings $g$, 
it reduces to an algebraic equation 
\begin{eqnarray}
 \Omega_{g \to \infty} 
 = (SS^\dagger)_{g \to \infty}  
 = c^{-1}H_0 e^{2My}H_0^\dagger 
  \label{SS-H0}
\end{eqnarray}
in the case of the infinite gauge coupling. 
In this limit our model reduces to the massive 
hyper-K\"ahler NLSM
on the cotangent bundle over the complex Grassmann manifold, 
$T^* G_{N_{\rm F},N_{\rm C}} = T^*[SU(N_{\rm F})/SU(N_{\rm C}) 
\times SU(N_{\rm F}-N_{\rm C})]$~\cite{ANS}. 
By choosing a gauge, we obtain uniquely the 
$N_{\rm C}\times N_{\rm C}$ complex matrix $S$ 
from the $N_{\rm C}\times N_{\rm C}$ Hermitian 
matrix $\Omega $. 
Then, we find that 
with a given arbitrary moduli matrix $H_0(z)$, 
explicit solutions for all the quantities, 
$\Sigma ,\,W_m$ and $H^1$ 
are obtained by Eqs.~(\ref{def-S-3}), ({\ref{def-S-bar}}) 
and (\ref{sol-H}). 
Therefore 
we can explicitly construct all solutions of the 1/4 BPS 
eqs.~(\ref{1/4BPSeq-1})-(\ref{1/4BPSeq-5})  exactly 
in the infinite gauge coupling. 

Our explicit solution shows that 
the total moduli space, including all topological sectors,  
is fully covered by our moduli matrix $H_0(z)$. 
Eqs.~(\ref{def-S-3}), (\ref{def-S-bar}) and (\ref{sol-H}) 
show that a left-multiplication to 
$S(x^m)$ and $H_0(z)$ by an arbitrary 
$V(z)$ of $GL(N_{\rm C},{\mathbf C})$, 
whose elements are holomorphic functions, 
give identical 
physical quantities $\Sigma ,\,W_m$ and $H^1$.
Since holomorphy of $H_0(z)$ must be respected, 
det$V(z)$ should be free of zeroes and 
singularities except at infinity. 
Therefore we find that 
the complete moduli space for solutions 
of the 1/4 BPS 
eqs.~(\ref{1/4BPSeq-1})-(\ref{1/4BPSeq-5}) 
is a set of whole 
holomorphic maps from the complex plane to the complex 
Grassmann manifold 
$ G_{N_{\rm F},N_{\rm C}}=\{H_0|H_0\simeq 
VH_0,\, V\in GL(N_{\rm C},{\mathbf C})\}$.
This result can be understood by noting that 
we obtain for each $z$ non-Abelian multi-wall solutions 
whose moduli space is $G_{N_{\rm F},N_{\rm C}}$~\cite{INOS}, 
while our 1/4 BPS solution may be regarded 
as a fully developed configuration of 
$z$-dependent ``fluctuations'' of moduli fields on walls.
Vortices reduce to NLSM lumps~\cite{SI} 
in $g^2\rightarrow \infty$~\cite{Sh,HT}. 
When spatial infinities of $z$ are mapped into 
a single point in $G_{N_{\rm F},N_{\rm C}}$, 
the $z$ plane can be compactified to ${\bf C}P^1$. 
Then the moduli space is the whole holomorphic maps from 
$S^2 \simeq {\bf C}P^1$ to $G_{N_{\rm F},N_{\rm C}}$, 
and the winding number is measured by 
$\pi_2 (G_{N_{\rm F},N_{\rm C}}) = {\bf Z}$  
with vortices winding the 2-cycles 
in $G_{N_{\rm F},N_{\rm C}}$. 
A crucial difference with ordinary lumps 
is 
that $G_{N_{\rm F},N_{\rm C}}$ is not 
the target space of a NLSM 
but the wall moduli space.

\begin{figure}[htb]
\includegraphics[width=8cm]{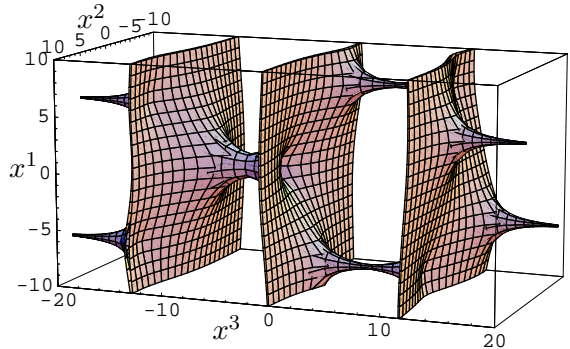}
\put(-140,5){$x^3$}
\put(-215,65){$x^1$}
\put(-190,120){$x^2$}
\caption{\label{fig:amidakuji} Multi-vorices 
between multi-walls: 
Surfaces defined by the same energy density 
$t_{\rm w}+t_{\rm v}=0.5c$ with 
$H_0(z)e^{Mx^3}=
\sqrt{c}((z-4-2i)(z+5+8i)e^{3/2x^3},
(z+8-i)(z-7+6i)e^{1/2x^3+12},
z^2e^{-1/2x^3+12},(z-6-5i)(z+6-7i)e^{-3/2x^3})$. }
\end{figure}

Our construction produces rich contents, 
even if we concentrate on 
the Abelian case ($N_{\rm C}=1$), 
which reduces to the massive NLSM on
$T^*{\bf C}P^{N_{\rm F}-1}$ in 
the strong coupling limit.
First consider the case where 
infinities of $z$ are mapped into 
a single point in $G_{N_{\rm F},N_{\rm C}}$. 
The quantity $\Omega $ reduces to a scalar 
\begin{eqnarray}
 \Omega =\sum_{A=1}^{N_{\rm F}}|f^A(z)|^2e^{2m_Ax^3}
\end{eqnarray}
with the moduli vector 
$H_0(z)=\sqrt{c}\left(f^1(z),\,\dots,
f^{N_{\rm F}}(z)\right)$. 
We have $N_{\rm F}$ vacua, 
which are ordered by the flavor label $A$, 
and maximally $N_{\rm F}-1$ parallel walls interpolating
between these vacua. 
For each fixed $z$, we can have maximally $N_{\rm F}-1$ 
walls at various points in $x^3$. 
By examining energy density, for instance, 
one can show the $\Omega $
describes a configuration close to the $A$-th vacuum, 
if only the $A$-th flavor is dominant in $\Omega$. 
If $A$-th and ($A+1$)-th flavors are comparable and dominant, it describes 
the $A$-th wall separating the vacua $A$ and $A+1$. 
The position 
of the $A$-th wall is 
easily guessed by comparing two 
adjacent flavors as 
$
x_A^3(z)=\left(\log|f_{A+1}(z)|-\log|f_A(z)|\right)/(m_A-m_{A+1})
$. 
The energy of the wall interpolating 
between the $A$-th and $B$-th vacua is given by 
\begin{eqnarray}
 \int^\infty_{-\infty} dx^3t_{\rm w}={c\over 2}
 \Big[\partial _3\log\Omega \Big]^\infty _{-\infty }=c(m_A-m_B). 
\end{eqnarray}
We find that 
walls are bent unless $f^A(z)$ is constant.    
Especially, if $f^A(z)$ has zeroes, 
walls are bent drastically and form vortices at 
those points. 
Actually, our solutions contain 
vortices stretched between walls at arbitrary 
positions.  
\begin{figure}[htb]
\includegraphics[width=4cm]{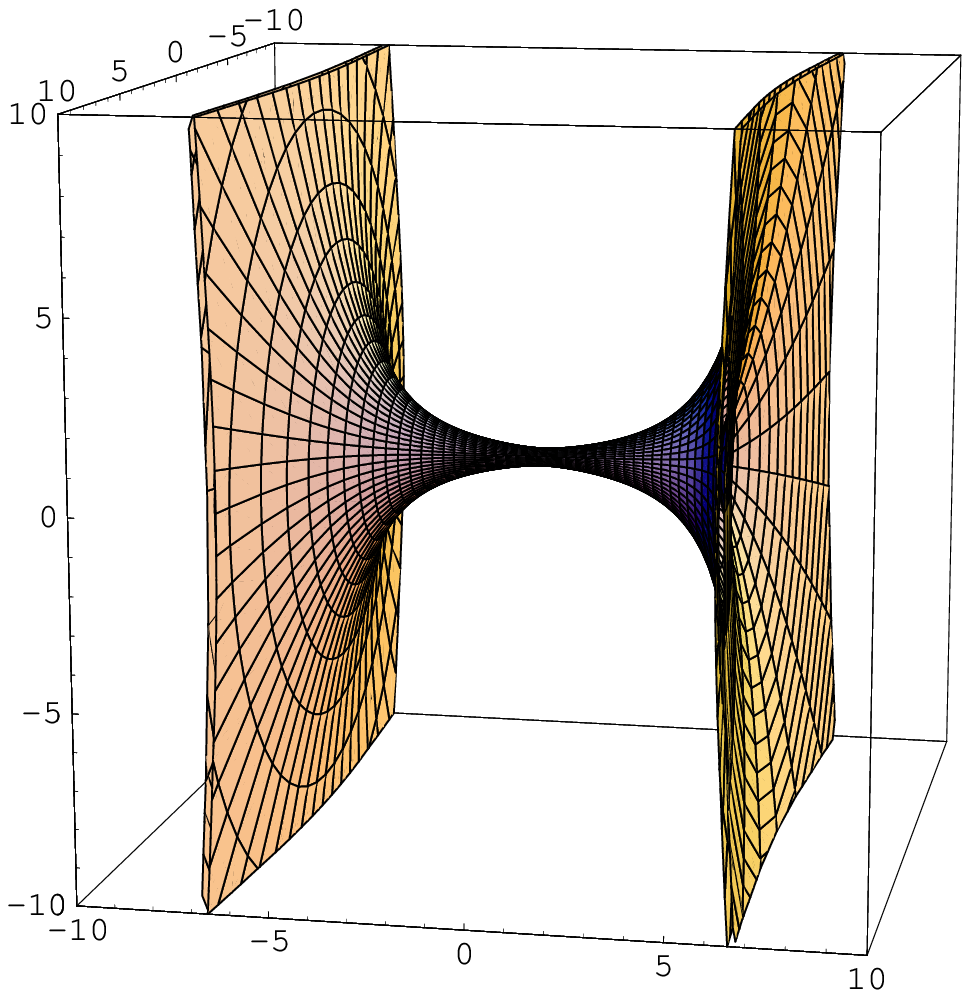}
\put(-75,-2){$x^3$}
\put(-115,60){$x^1$}
\put(-100,110){$x^2$}
\put(-60,-7){a)}
\,
\includegraphics[width=4cm]{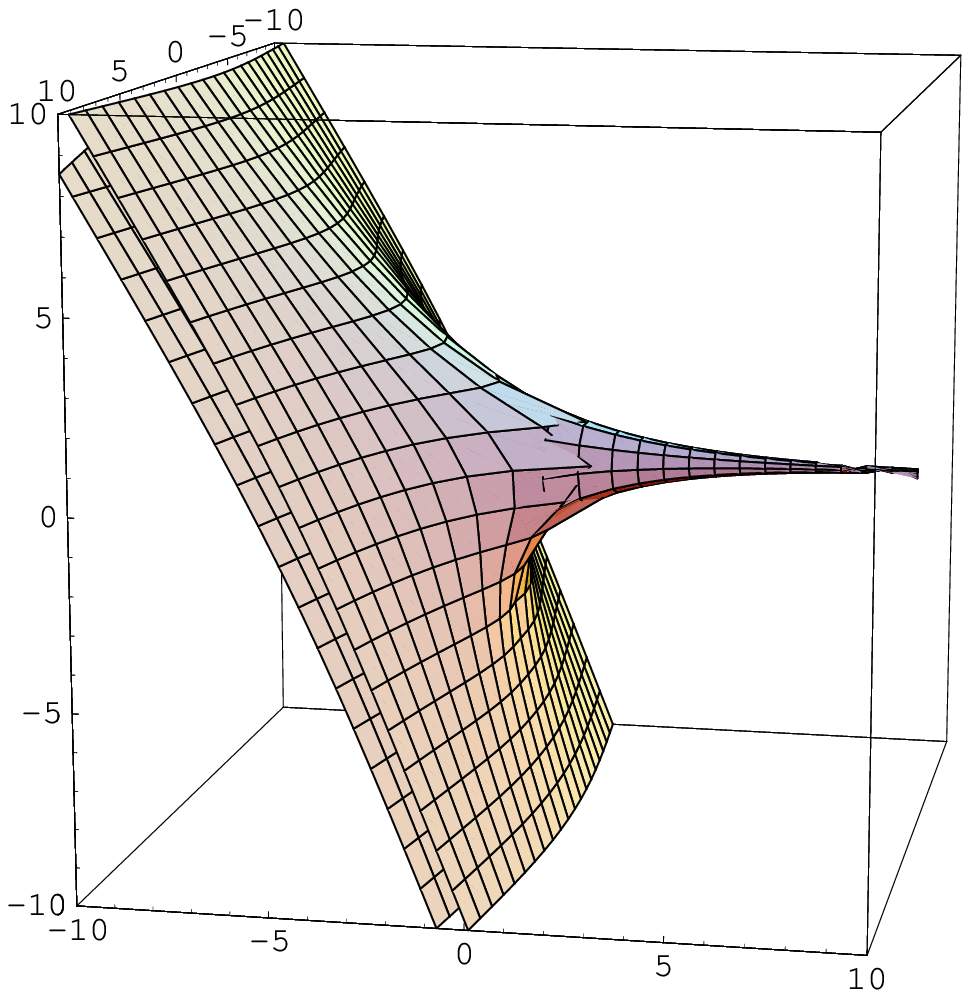}
\put(-75,-2){$x^3$}
\put(-115,60){$x^1$}
\put(-100,110){$x^2$}
\put(-60,-7){b)}
\caption{\label{fig:wormhole} Surfaces defined by the same energy density 
$t_{\rm w}+t_{\rm v}=0.5c$: 
a) A vortex stretched between walls with 
$H_0(z)e^{Mx^3}=\sqrt{c}(e^{x^3},ze^{4},e^{-x^3})$. 
b) A vortex 
attached to a tilted wall with 
$H_0(z)e^{Mx^3}=\sqrt{c}(z^2e^{x^3},e^{-1/2z})$.
Note that there are two surfaces with the same energy 
for each wall.}
\end{figure}
To see this let us choose 
the moduli matrix 
\begin{eqnarray}
 f^A(z)=f^A_0\prod_{\alpha }(z-z^A_{\alpha })^{k^A_{\alpha }}, 
  \quad k^A_\alpha \in {\bf Z}_+, 
\end{eqnarray}
which gives vortices of vorticity $k_{\alpha}^A$ 
at $z=z^A_{\alpha}$ in the $A$-th vacuum. 
Fig.\ref{fig:wormhole} a) illustrates a vortex stretching 
between two walls, where logarithmic bending of the wall 
 is visible towards $|z|\rightarrow \infty $. 
To avoid the logarithmic bending, 
we require 
$ k=\sum_{\alpha }k^A_\alpha$ to be common to walls, 
as shown in Fig.\ref{fig:amidakuji}. 
We 
obtain the vorticity by an integration on a disc $D$ 
with infinite radius 
\begin{eqnarray}
 \int _Dd^2x{t_{\rm v} \over 2\pi c}
=
-i{1 \over 2\pi} \int _{\partial D}dz\partial (\log\Omega )
= k
\end{eqnarray}
using $\Omega \propto |z|^{2k}$ at $|z|\rightarrow \infty $.
In the effective field theory on the D-brane, 
a brane ending on a single brane has been obtained, 
but a complete solution of 
branes stretching between 
two or more 
branes
was difficult to achieve~\cite{BIon}. 
Our construction generalizes D-brane 
soliton in \cite{GPTT},
and may give insight into string dynamics.

A monopole in the Higgs phase 
was found recently in 
non-Abelian gauge theories~\cite{mono-Higgs}. 
We will now show 
that a similar monopole in the Higgs phase also exists in 
$U(1)$ gauge theory. 
Because of $1/g^2$ factor, 
the energy density of monopoles $t_{\rm m}$ vanish 
in the limit of infinite gauge coupling. 
The monopole charge $g^2t_{\rm m}$ is, however, 
finite as a kink on the vortex, precisely analogous 
to the non-Abelian case~\cite{mono-Higgs}. 
In a simple example of a vortex with winding number $k$ 
stretched between two walls, we obtain 
the monopole charge 
\begin{eqnarray}
\int _Vd^3x g^2 t_{\rm m}=-\pi |m_A-m_B|k.
\end{eqnarray}
Let us stress that our monopole in the Higgs phase should 
give non-vanishing contribution to the energy density once 
gauge coupling becomes finite.

In the case where 
infinities of $z$ are mapped into 
a single point in $G_{N_{\rm F},N_{\rm C}}$, 
walls are perpendicular to, 
and vortices are extending 
along, the $x^3$-axis in our 1/4 BPS solutions. 
If we relax this condition and 
allow an exponential function of $z$, 
such as $e^{m z}$, 
somewhere in the moduli matrix $H_0(z)$, 
the corresponding wall is no longer perpendicular to 
the $x^3$-axis. 
If we choose $H_0(z)$, for instance, 
\begin{eqnarray}
H_0e^{Mx^3}\!=\! 
\sqrt{c}(e^{m_1x^3+\tilde m_1 z},\,e^{m_2x^3+\tilde m_2 z}),
\,\,\tilde m_{1, 2} \in {\mathbf C}, 
\end{eqnarray}
we can guess that the wall position is expressed as
$m_1x^3+{\rm Re}(\tilde m_1 z)=$ $m_2x^3+{\rm Re}(\tilde m_2 z)$. 
Actually we find the energy density 
$
 t_{\rm wv}=(c/2)\left(\Delta m^2+|\Delta \tilde m|^2\right)
{\rm sech}^2(X) 
$ 
to depend only on 
$X\equiv \Delta m x^3+{\rm Re}(\Delta \tilde m z) $ 
with $\Delta m=m_1-m_2$ and 
$\Delta \tilde m=\tilde m_1-\tilde m_2$. 
We thus find that the wall configuration 
is perpendicular to a vector 
$(\Delta m,\,{\rm Re}(\Delta \tilde m),\,-{\rm Im}(\Delta \tilde m))$. 
Moreover we find 
the dual field strength 
\begin{eqnarray}
{}^*F_3&=&-2\partial \bar \partial \log\Omega 
=-{1\over 2}{|\Delta \tilde m|^2{\rm sech}^2(X)},\nonumber\\
 {}^*F_1+i{}^*F_2&=&\bar\partial \partial _3\log\Omega 
={1\over 2}{\Delta m \Delta \tilde m^*{\rm sech}^2(X)},
\end{eqnarray}  
flows down along the tilted wall 
to negative infinity of $x^3$: 
$\Delta m F_3+\Delta \tilde m 
\left({}^*F_1+i{}^*F_2\right)=0$. 
If vortices are present, they are no longer perpendicular 
to such tilted walls as illustrated 
in Fig.~\ref{fig:wormhole} b).
This configuration offers a field theoretical model 
of the string ending on the D-brane with a magnetic 
flux~\cite{HH}.

\begin{figure}[htb]
\includegraphics[width=6.5cm]{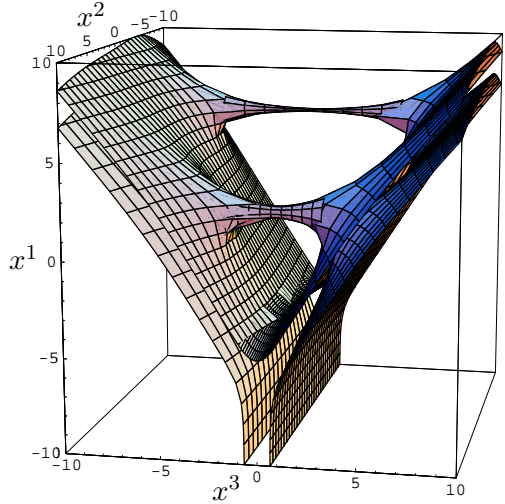}
\put(-110,0){$x^3$}
\put(-185,85){$x^1$}
\put(-160,175){$x^2$}
\caption{\label{fig:ayatori} A cat's-cradle 
soliton: 
Surfaces defined by the same energy density 
$t_{\rm w}+t_{\rm v}=0.5c$ with 
$H_0(z)e^{Mx^3}=
\sqrt{c}(e^{x^3},(z-2-5i)(z-6+5i)e^{3/4z-1/2},e^{-x^3})$. }
\end{figure}

We can construct a domain wall junction 
from two tilted walls 
using $H_0(z)$ 
with $\tilde m_1,\,\tilde m_2,\,\tilde m_3\in {\mathbf C}$ 
\begin{eqnarray}
 H_0(z)=\sqrt{c}(e^{\tilde m_{1}z},\,e^{\tilde m_2 z},\,
e^{\tilde m_{3}z}).
\end{eqnarray}
Positions $x^3_1(z),\,x^3_2(z)$ of the two walls can 
be guessed as  
$
 x^3_A(z)
={\rm Re}((\tilde m_{A+1} -\tilde m_A)z)/( m_A-m_{A+1})$, 
for $A=1,2$, 
which is a good estimate when $x^3_1(z)\gg x^3_2(z)$. 
In regions where $x^3_1(z) < x^3_2(z)$, 
however, we find that 
the configuration describes just a single wall whose position 
is given by the center of mass 
$
{{\rm Re}((\tilde m_{3} -\tilde m_1)z)}/{(m_1-m_{3})}
$. 
Therefore the solution gives a junction of three walls 
which meet at $x^3=x^3_1(z)=x^3_2(z)$. 
These features are visible in Fig.\ref{fig:ayatori} 
where we show a ``cat's-cradle'' soliton 
as a complicated example of composite solitons which can 
be easily constructed as an exact solution by our method. 

We have constructed composite solitons exactly 
in contrast to a description by 
the effective field theory on the host brane~\cite{BIon},  
which is valid as an approximation for small fluctuation.
Our method gives all possible solutions  exactly 
also for non-Abelian $U(N_{\rm C})$ case easily. 
We have used $U(1)$ case merely to illustrate the power 
of our method. 

\begin{acknowledgments}
We thank Koji Hoshimoto for a useful discussion. 
This work is supported in part by Grant-in-Aid for Scientific 
Research from the Ministry of Education, Culture, Sports, 
Science and Technology, Japan No.13640269 
and 16028203 (NS and MN). 
K.O.~and M.N.~are 
supported in part by 
JSPS and 
Y.I.~a 21st Century COE Program at 
Tokyo Tech ``Nanometer-Scale Quantum Physics''. 
\end{acknowledgments}

\bibliography{}
\newcommand{\J}[4]{{\sl #1} {\bf #2} (#3) #4}
\newcommand{\andJ}[3]{{\bf #1} (#2) #3}
\newcommand{\AP}{Ann.\ Phys.\ (N.Y.)}
\newcommand{\MPL}{Mod.\ Phys.\ Lett.}
\newcommand{\NP}{Nucl.\ Phys.}
\newcommand{\PL}{Phys.\ Lett.}
\newcommand{\PR}{ Phys.\ Rev.}
\newcommand{\PRL}{Phys.\ Rev.\ Lett.}
\newcommand{\PTP}{Prog.\ Theor.\ Phys.}
\newcommand{\hep}[1]{{\tt hep-th/{#1}}}

\end{document}